\documentclass[12pt]{article}
\usepackage[latin9]{inputenc}
\usepackage[a4paper]{geometry}
\usepackage{booktabs}
\usepackage{mathtools}
\usepackage{amsmath}
\usepackage{amsthm}
\usepackage{amssymb}
\usepackage{setspace}
\usepackage[authoryear]{natbib}
\makeatletter

\providecommand{\tabularnewline}{\\}

\usepackage{amsthm}

\providecommand{\tabularnewline}{\\}

\theoremstyle{plain}
\theoremstyle{definition}
\theoremstyle{plain}

\newtheorem{assumption}{Assumption}

\providecommand{\examplename}{Example}
\providecommand{\propositionname}{Proposition}
\providecommand{\theoremname}{Theorem}

\renewcommand{\hat}{\widehat}

\makeatother

\begin{document}
\title{Two-Sample IV: Efficient Two-Step Estimation and Tests for Overidentification
and Weak-Instruments}
\author{Fatima Kasenally, Ruoxi Guan and Frank Windmeijer \\
 {\small Dept of Statistics, University of Oxford, UK}\\
 }
\date{13 February, 2026}
\maketitle
\begin{abstract}
\noindent\baselineskip=15pt Two-sample IV is a popular estimation
method when the outcome and treatment variables are available in different
samples, whereas instruments are available in both samples. The standard
estimator is two-sample two-stage least squares estimator, which is
efficient under homoskedasticity and homogeneity of the samples. We
develop a robust two-step procedure for efficient estimation under
general heteroskedasticity and heterogeneity of the samples, and propose
a related two-sample Hansen overidentification test. A key feature
of our approach is that only summary statistics from the linear regressions
of the reduced form and first-stage in the two samples are needed.
These are the six objects of the estimated coefficient vectors, and
the homoskedastic and heteroskedasticity robust estimated variance
matrices. We further show that the first-stage $F$-statistic in the
treatment sample can be used as a test for weak instruments in the
standard way under homoskedasticity and homogeneity, with the relative
bias here a proportional bias. We propose an extension of the effective
$F$-statistic of \citet{Olea2013} for the heteroskedastic case,
following the generalization in \citet{Windmeijer2025}. We illustrate
the estimators and tests in an application studying the effect of
education on voting behavior from \citet{marshall2019anti}, with
cluster robust inference.
\end{abstract}

\section{Introduction}

Two-sample instrumental variables estimation was developed by \citet{Klevmarken1982},
\citet[1995]{AngristKruegerJASA1992}, \nocite{AngristKruegerJBES1995}
\citet{ArellanoMeghirREStud1992} and \citet[2010]{InoueSolonWP2005}\nocite{InoueSolonREStat2010},
among others. It applies to settings where an outcome $y$ and possibly
endogenous explanatory variables $x$ are not observed in the same
data set. Instead, in the standard two-sample setup we consider here,
one has observations on $y$, instruments $z$ and covariates $c$
in one sample and observations on $x$, $z$ and $c$ in another sample.

The commonly used estimator in the linear setting we consider is the
two-sample two-stage least squares (ts2sls) estimator as introduced
by \citet{Klevmarken1982}. This estimator is asymptotically normally
distributed and efficient under standard sampling, homoskedasticity
and homogeneity assumptions of the distributions of the two samples.

In the standard one-sample IV setting, when there are more instruments
than endogenous variables and so the specification is overidentified,
test results for overidentification are routinely reported. Standard tests are the Sargan test, which is valid under homoskedasticity, and the robust
Hansen $J$-test, where robustness is to general forms of heteroskedasticity,
including for example clustering. The $J$-test statistic is calculated
using a two-step estimation procedure, where the two-step estimator
is efficient under heteroskedasticity. Such a two-step estimator has
not been considered for the two-sample setup and we develop it here,
resulting in the two-sample $J$-test. A prime feature of our derivations
is that we only need the summary statistics from two linear regressions,
that of $y$ on $z$ and $c$ in sample 1 and $x$ on $z$ and $c$
in sample 2. These summary statistics are the vectors of estimated
coefficients, their estimated variance matrices valid under homoskedasticity
and the heteroskedasticity robust variance estimators. From these
six objects we obtain the ts2sls estimator, their standard errors,
including the robust ones as in \citet{PaciniWindEL2016}, the efficient
two-step estimator and the $J$-test statistic. We derive the (standard)
limiting distributions fully, obtained simply from those of the least
squares estimators in the two samples. This is done in Sections \ref{sec:Model}
and \ref{sec:Two-step-Estimation}, with Section \ref{sec:Some-Simulation-Results}
providing some simulation results.

The paper closest to our setup is \citet*{ZhaoetalStatSci2019}. They
did consider the test for overidentifying restrictions, but in a homoskedastic
setting, allowing for heterogeneity of the two samples. They did not
fully establish the limiting distribution results, which we do also
for their setting. Another related paper is \citet*{ChoietalJAE2017},
who propose the two-sample robust Anderson-Rubin statistic for weak-instrument
robust inference, but do not consider the efficient two-step estimator
and the associated two-sample $J$-test.

We further consider testing for weak instruments in Section \ref{sec:Weak-Instruments-Tests,-Bias}.
For the one-variable, homoskedasticity and homogeneous samples case,
we derive the relationship between the first-stage $F$-statistic
and the relative bias of the ts2sls estimator of $\beta$ under weak-instrument
asymptotics as in \citet{Staiger1997} and \citet{Stock2005}. For
the ts2sls estimator, the weak-instruments bias is towards $0$ and
we find that the critical values of \citet{Stock2005} apply for the
the relative bias results, where here the relative bias is relative
to the true parameter value $\beta$.

\citet{Olea2013} propose the effective $F$-statistic for testing
weak instruments for the standard one-sample 2sls estimator under
heteroskedasticity. They provide critical value functions for the
effective $F$-statistic for the Nagar bias of the 2sls estimator,
relative to a benchmark bias. \citet{Windmeijer2025} generalizes
the effective $F$-statistic to a general GMM setting, and we apply
his results to derive the effective $F$-statistic as a test for weak
instruments, related to the Nagar bias of the ts2sls estimator again
relative to the true parameter value $\beta$.

Section \ref{sec:Application} presents estimation and test results
for the study of \citet{marshall2019anti} of the effects of education
on political affiliation, a two-sample IV analysis wit cluster-robust
inference.

\section{Model}

\label{sec:Model}

For ease of exposition, we focus here an a single explanatory variable
$x$, the details for multivariable $x$ are presented in the Appendix.
We have the model specifications 
\begin{align*}
y & =x\beta+u\\
x & =z'\pi_{x}+v,
\end{align*}
where other exogenous explanatory variables that are observed in both
data sets have been linearly partialled out. $z$ and $\pi_{x}$ are
$k_{z}$-vectors.

The reduced form for $y$ is then given by 
\begin{align*}
y & =z'\pi_{x}\beta+u+v\beta\\
 & =z'\pi_{y}+v_{y}
\end{align*}
with the $k_{z}$-vector $\pi_{y}=\pi_{x}\beta$ and $v_{y}=u+v\beta$.

We don't observe $y$, $x$ and $z$ in one sample, but $y$ and $z$
in sample 1, $x$ and $z$ in sample 2, the observations being $\left\{ y_{1i},z_{1i}^{\prime}\right\} _{i=1}^{n_{1}}$
and $\left\{ x_{2j},z_{2j}^{\prime}\right\} _{j=1}^{n_{2}}$. The
samples are independent and we assume that $\lim_{n_{1}\rightarrow\infty,n_{2}\rightarrow\infty}\frac{n_{1}}{n_{2}}=\alpha$,
with $\alpha>0$ and finite. As in \citet{ZhaoetalStatSci2019} we
make the following structural assumption for the two samples.

\begin{assumption}\label{ass:struc} The two samples satisfy the
reduced form specifications 
\begin{align*}
y_{1i} & =z_{1i}^{\prime}\pi_{y}+v_{y_{1},i}\\
x_{2j} & =z_{2j}^{\prime}\pi_{x}+v_{x_{2},i};
\end{align*}
\[
\mathbb{E}\left(z_{2j}v_{x_{2},j}\right)=0;
\]
\[
\pi_{y}=\pi_{x}\beta.
\]

\end{assumption}

We further assume that the instruments are relevant and valid.

\begin{assumption}\label{ass:relval}

The instruments are relevant, $\pi_{x}\neq0$, and valid, $\mathbb{E}\left(z_{1i}v_{y_{1},i}\right)=0$.

\end{assumption}

As in \citet{ZhaoetalStatSci2019}, under Assumption
\ref{ass:struc} together with Assumption \ref{ass:relval}, we obtain
all our results for estimation and testing allowing for heterogenous
samples. The homogeneity assumption of the samples is formulated as
follows, see

\begin{assumption} \label{ass:homg}Homogeneous samples. The following
moments are the same in the two samples. 
\begin{align*}
\mathbb{E}\left(z_{1i}z_{1i}^{\prime}\right) & =\mathbb{E}\left(z_{2j}z_{2j}^{\prime}\right);\\
\mathbb{E}\left(z_{1i}x_{1i}^{\prime}\right) & =\mathbb{E}\left(z_{2j}x_{2j}^{\prime}\right).
\end{align*}

\end{assumption}

For homogeneous samples, it follows that the first-stage relationship
$x=z'\pi_{x}+v$ does not need to be structural, but can be a linear
projection. It follows from Assumption \ref{ass:homg} that the linear
projection parameters in the two samples are the same and equal to
$\pi_{x}$. It follows then that $\pi_{y}=\pi_{x}\beta$ and from
the linear projection that $\mathbb{E}\left(z_{2j}v_{x_{2},j}\right)=0$.

As mentioned, for the results obtained below we assume Assumptions
\ref{ass:struc} and \ref{ass:relval}. Some results simplify when
the homogeneity Assumption \ref{ass:homg} holds, which we will highlight
specifically. For example, for using the standard $F$-statistic as
a test for weak-instruments we need the homogeneity assumption.

Let $y_{1}$ and $x_{2}$ be the $n_{1}$- and $n_{2}$-vectors $\left(y_{1i}\right)$
and $\left(x_{2j}\right)$ respectively, and $Z_{1}$ and $Z_{2}$
the $\left(n_{1}\times k_{z}\right)$ and $\left(n_{2}\times k_{z}\right)$
matrices of observations of the instruments in the two samples. The
ts2sls estimator of $\beta$ is obtained as the OLS estimator for
$\beta$ in the specification 
\begin{align}
y_{1} & =\hat{x}_{1}\beta+\varepsilon;\label{eq:xhat}\\
\hat{x}_{1} & =Z_{1}\hat{\pi}_{x,2},\nonumber 
\end{align}
resulting in 
\[
\hat{\beta}_{ts2sls}=\left(\hat{x}_{1}^{\prime}\hat{x}_{1}\right)^{-1}\hat{x}_{1}^{\prime}y_{1}=\left(\hat{\pi}_{x,2}^{\prime}Z_{1}^{\prime}Z_{1}\hat{\pi}_{x,2}\right)^{-1}\hat{\pi}_{x,2}^{\prime}Z_{1}^{\prime}Z_{1}\hat{\pi}_{y,1},
\]
where $\hat{\pi}_{x,2}$ and $\hat{\pi}_{y,1}$ are the OLS estimators
of $\pi_{x}$ and $\pi_{y}$ given by 
\begin{align*}
\hat{\pi}_{y,1} & =\left(Z_{1}^{\prime}Z_{1}\right)^{-1}Z_{1}^{\prime}y_{1}\\
\hat{\pi}_{x,2} & =\left(Z_{2}^{\prime}Z_{2}\right)^{-1}Z_{2}^{\prime}x_{2}.
\end{align*}

Under Assumption 1 and standard regularity conditions for OLS estimators,
the limiting distributions are given by 
\begin{align}
\sqrt{n_{1}}\left(\hat{\pi}_{y,1}-\pi_{y}\right) & \stackrel{d}{\rightarrow}\mathcal{N}\left(0,\Sigma_{\hat{\pi}_{y,1}}\right)\label{eq:limy1}\\
\sqrt{n_{2}}\left(\hat{\pi}_{x,2}-\pi_{x}\right) & \stackrel{d}{\rightarrow}\mathcal{N}\left(0,\Sigma_{\hat{\pi}_{x,2}}\right).\label{eq:limx2}
\end{align}
For the estimators of the variance of $\hat{\pi}_{y,1}$ and $\hat{\pi}_{x,2}$,
we distinguish between those that are valid under conditional homoskedasticity
and those that are robust to general forms of heteroskedasticity,
including serial correlation and clustering. For example, the homoskedastic
variance estimator for $\hat{\pi}_{y,1}$ is given by 
\[
\hat{V}_{\hat{\pi}_{y,1}}=\hat{\sigma}_{v_{y,1}}^{2}\left(Z_{1}^{\prime}Z_{1}\right)^{-1},
\]
where $\hat{\sigma}_{v_{y,1}}^{2}=\frac{1}{n}\hat{v}_{y,1}^{T}\hat{v}_{y,1}$,
with $\hat{v}_{y,1}=y_{1}-Z_{1}\hat{\pi}_{y,1}$. $\hat{V}_{\hat{\pi}_{y,1}}$
is an estimator of $\Sigma_{\hat{\pi}_{y,1}}/n_{1}$ in the sense
that $n_{1}\hat{V}\left(\hat{\pi}_{y,1}\right)\stackrel{p}{\rightarrow}\Sigma_{\hat{\pi}_{y,1}}$
if $\mathbb{E}\left(v_{y,1i}^{2}|z_{1i}\right)=\sigma_{v_{y_{1}}}^{2}$.

As an example, a robust variance estimator in a cross-section with
conditionally heteroskedastic errors is given by 
\[
\hat{V}_{r,\hat{\pi}_{y,1}}=\left(Z_{1}^{\prime}Z_{1}\right)^{-1}\left(\sum_{i=1}^{n_{1}}\hat{v}_{y,1}^{2}z_{i}z_{i}^{\prime}\right)\left(Z_{1}^{\prime}Z_{1}\right)^{-1}.
\]

From the two OLS regressions, we collect the six objects 
\[
\left\{ \hat{\pi}_{y,1},\hat{V}_{\hat{\pi}_{y,1}},\hat{V}_{r,\hat{\pi}_{y,1}},\hat{\pi}_{x,2},\hat{V}_{\hat{\pi}_{x,2}},\hat{V}_{r,\hat{\pi}_{x,2}}\right\} .
\]
These six objects are the only ones needed to obtain the ts2sls estimator,
its (robust) variance estimator, the efficient two-step estimator
and two-sample Sargan and $J$-test statistics.

First, consider the one-step GMM-type estimator 
\begin{align*}
\hat{\beta}_{1s} & =\arg\min_{b}\left(\hat{\pi}_{y,1}-\hat{\pi}_{x,2}b\right)^{\prime}\hat{V}_{\hat{\pi}_{y,1}}^{-1}\left(\hat{\pi}_{y,1}-\hat{\pi}_{x,2}b\right)\\
 & =\left(\hat{\pi}_{x,2}^{\prime}\hat{V}_{\hat{\pi}_{y,1}}^{-1}\hat{\pi}_{x,2}\right)^{-1}\hat{\pi}_{x,2}^{T}\hat{V}_{\hat{\pi}_{y,1}}^{-1}\hat{\pi}_{y,1}\\
 & =\hat{\beta}_{ts2sls}.
\end{align*}
From the relationship $\pi_{y}=\pi_{x}\beta$, it follows that 
\begin{equation}
\hat{\pi}_{y,1}=\hat{\pi}_{x,2}\beta+\left(\hat{\pi}_{y,1}-\pi_{y}\right)-\left(\hat{\pi}_{x,2}-\pi_{x}\right)\beta.\label{eq:gamhatexp}
\end{equation}
Relationship (\ref{eq:gamhatexp}) is a crucial observation that facilitates
obtaining the limiting distribution of the ts2sls estimator from the
limiting distributions of the OLS estimators as given in (\ref{eq:limy1})
and (\ref{eq:limx2}), and which was missing from the analysis in
\citet{ZhaoetalStatSci2019}. It follows from (\ref{eq:gamhatexp})
that 
\[
\sqrt{n_{1}}\left(\hat{\beta}_{ts2sls}-\beta\right)=\left(\hat{\pi}_{x,2}^{T}\hat{V}_{\hat{\pi}_{y,1}}^{-1}\hat{\pi}_{x,2}\right)^{-1}\hat{\pi}_{x,2}^{T}\hat{V}_{\hat{\pi}_{y,1}}^{-1}\sqrt{n_{1}}\left(\left(\hat{\pi}_{y,1}-\pi_{y}\right)-\left(\hat{\pi}_{x,2}-\pi_{x}\right)\beta\right)
\]
and so from the limiting distributions (\ref{eq:limy1}) and (\ref{eq:limx2})
and independence of the two samples, it follows that 
\[
\sqrt{n_{1}}\left(\hat{\beta}_{ts2sls}-\beta\right)\stackrel{d}{\rightarrow}\mathcal{N}\left(0,\Sigma_{\hat{\beta}_{ts2sls}}\right)
\]
where 
\[
\Sigma_{\hat{\beta}_{ts2sls}}=C\left(\pi_{x},\Sigma_{\hat{\pi}_{y,1}}^{-1}\right)^{T}\left(\Sigma_{\hat{\pi}_{y,1}}+\beta^{2}\alpha\Sigma_{\hat{\pi}_{x,2}}\right)C\left(\pi_{x},\Sigma_{\hat{\pi}_{y,1}}^{-1}\right),
\]
with 
\[
C\left(\pi_{x},\Sigma_{\hat{\pi}_{y,1}}^{-1}\right)\coloneqq\Sigma_{\hat{\pi}_{y,1}}^{-1}\pi_{x}\left(\pi_{x}^{\prime}\Sigma_{\hat{\pi}_{y,1}}^{-1}\pi_{x}\right)^{-1}.
\]

It follows that, under homoskedasticity, the variance of $\hat{\beta}_{ts2sls}$
can be estimated by 
\[
\hat{V}_{\hat{\beta}_{ts2sls}}=C\left(\hat{\pi}_{x,2},\hat{V}_{\hat{\pi}_{y,1}}^{-1}\right)'\left(\hat{V}_{\hat{\pi}_{y,1}}+\hat{\beta}_{ts2sls}^{2}\hat{V}_{\hat{\pi}_{x,2}}\right)C\left(\hat{\pi}_{x,2},\hat{V}_{\hat{\pi}_{y,1}}^{-1}\right),
\]
whereas under heteroskedasticity, the robust variance estimator is
given by 
\[
\hat{V}_{r,\hat{\beta}_{ts2sls}}=C\left(\hat{\pi}_{x,2},\hat{V}_{\hat{\pi}_{y,1}}^{-1}\right)'\left(\hat{V}_{r,\hat{\pi}_{y,1}}+\hat{\beta}_{ts2sls}^{2}\hat{V}_{r,\hat{\pi}_{x,2}}\right)C\left(\hat{\pi}_{x,2},\hat{V}_{\hat{\pi}_{y,1}}^{-1}\right),
\]
which is the result of \citet{PaciniWindEL2016}.

\section{Two-step Estimation and Overidentification Test}

\label{sec:Two-step-Estimation}

We now consider the case of general heteroskedasticity and efficient
two-step estimation. Let $\hat{\beta}_{ts2sls}$ be the initial, one-step
estimator as described above. Let 
\[
W_{n,r}\left(\hat{\beta}_{ts2sls}\right)=\left(\hat{V}_{r,\hat{\pi}_{y,1}}+\hat{\beta}_{ts2sls}^{2}\hat{V}_{r,\hat{\pi}_{x,2}}\right)^{-1},
\]
then the efficient two-step estimator is given by 
\[
\hat{\beta}_{2s}=\left(\hat{\pi}_{x,2}^{\prime}W_{n,r}\left(\hat{\beta}_{ts2sls}\right)\hat{\pi}_{x,2}\right)^{-1}\hat{\pi}_{x,2}^{\prime}W_{n,r}\left(\hat{\beta}_{ts2sls}\right)\hat{\pi}_{y,1},
\]
with limiting distribution 
\[
\sqrt{n_{1}}\left(\hat{\beta}_{2s}-\beta\right)\stackrel{d}{\rightarrow}\mathcal{N}\left(0,\left(\pi_{x}^{\prime}\left(\Sigma_{\hat{\pi}_{y,1}}+\beta^{2}\alpha\Sigma_{\hat{\pi}_{x,2}}\right)^{-1}\pi_{x}\right)^{-1}\right).
\]
An estimator for the variance of $\hat{\beta}_{2s}$ is therefore
given by 
\[
\hat{V}_{r,\hat{\beta}_{2s}}=\left(\hat{\pi}_{x,2}^{\prime}W_{n,r}\left(\hat{\beta}_{ts2sls}\right)\hat{\pi}_{x,2}\right)^{-1}.
\]
Efficiency of this two-step GMM estimator is a standard result.

The test for overidentifying restrictions then follows. Under the
null $H_{0}:\pi_{y}=\pi_{x}\beta$ and the standard assumptions we
have that 
\[
J\left(\hat{\beta}_{2s};\hat{\beta}_{ts2sls}\right)=\left(\hat{\pi}_{y,1}-\hat{\pi}_{x,2}\hat{\beta}_{2s}\right)^{T}W_{n,r}\left(\hat{\beta}_{ts2sls}\right)\left(\hat{\pi}_{y,1}-\hat{\pi}_{x,2}\hat{\beta}_{2s}\right)\stackrel{d}{\rightarrow}\chi_{k_{z}-1}^{2}.
\]
This is again a standard result, shown as follows. We have that 
\begin{align*}
\hat{\pi}_{y,1}-\hat{\pi}_{x,2}\hat{\beta}_{2s} & =\left(I_{k_{z}}-\hat{\pi}_{x,2}\left(\hat{\pi}_{x,2}^{\prime}W_{n,r}\left(\hat{\beta}_{ts2sls}\right)\hat{\pi}_{x,2}\right)^{-1}\hat{\pi}_{x,2}^{\prime}W_{n,r}\left(\hat{\beta}_{ts2sls}\right)\right)\hat{\pi}_{y,1}\\
 & =\left(I_{k_{z}}-\hat{\pi}_{x,2}\left(\hat{\pi}_{x,2}^{\prime}W_{n,r}\left(\hat{\beta}_{ts2sls}\right)\hat{\pi}_{x,2}\right)^{-1}\hat{\pi}_{x,2}^{\prime}W_{n,r}\left(\hat{\beta}_{ts2sls}\right)\right)\left(\hat{\pi}_{y,1}-\hat{\pi}_{x,2}\beta\right).
\end{align*}
It follows that 
\[
W_{n,r}^{1/2}\left(\hat{\beta}_{ts2sls}\right)\left(\hat{\pi}_{y,1}-\hat{\pi}_{x,2}\hat{\beta}_{2s}\right)
\]
\begin{align*}
 & =\left(I_{k_{z}}-W_{n,r}^{1/2}\left(\hat{\beta}_{ts2sls}\right)\hat{\pi}_{x,2}\left(\hat{\pi}_{x,2}^{T}W_{n,r}\left(\hat{\beta}_{ts2sls}\right)\hat{\pi}_{x,2}\right)^{-1}\hat{\pi}_{x,2}^{T}W_{n,r}^{1/2}\left(\hat{\beta}_{ts2sls}\right)\right)\\
 & \,\,\,\,\,\,\,\,\,\,\,\,\,\,\,\times W_{n,r}^{1/2}\left(\hat{\beta}_{ts2sls}\right)\left(\hat{\pi}_{y,1}-\hat{\pi}_{x,2}\beta\right)\\
 & \eqqcolon\left(I_{k_{z}}-A_{n}\right)W_{n,r}^{1/2}\left(\hat{\beta}_{ts2sls}\right)\left(\hat{\pi}_{y,1}-\hat{\pi}_{x,2}\beta\right),
\end{align*}
As 
\[
W_{n,r}^{1/2}\left(\hat{\beta}_{ts2sls}\right)\left(\hat{\pi}_{y,1}-\hat{\pi}_{x,2}\beta\right)\stackrel{d}{\rightarrow}\mathcal{N}\left(0,I_{k_{z}}\right)
\]
and $\left(I_{k_{z}}-A_{n}\right)$ is a symmetric idempotent matrix
with rank equal to $tr$$\left(I_{k_{z}}-A_{n}\right)=k_{z}-1$, the
result follows.

\subsection{Homoskedasticity}

Under Assumptions \ref{ass:struc} and \ref{ass:relval}, leaving out the homogeneity assumption and the assumption of conditional homoskedasticity,
we obtain the two-step estimator using the weight matrix based on
the homoskedastic variance estimator 
\[
W_{n}\left(\hat{\beta}_{ts2sls}\right)=\left(\hat{V}_{\hat{\pi}}+\hat{\beta}_{ts2sls}^{2}\hat{V}_{\hat{\pi}_{x,2}}\right)^{-1}.
\]
Denoting this estimator $\hat{\beta}_{2s,hom},$with the Sargan test
statistic then given by 
\[
S\left(\hat{\beta}_{2s,hom};\hat{\beta}_{ts2sls}\right)=\left(\hat{\pi}_{y,1}-\hat{\pi}_{x,2}\hat{\beta}_{2s,hom}\right)^{T}W_{n}\left(\hat{\beta}_{ts2sls}\right)\left(\hat{\pi}_{y,1}-\hat{\pi}_{x,2}\hat{\beta}_{2s,hom}\right)\stackrel{d}{\rightarrow}\chi_{k_{z}-1}^{2}.
\]

Under homogeneity Assumption \ref{ass:homg}, we don't need a two-step
estimator, as then 
\[
S\left(\hat{\beta}_{ts2sls}\right)=\left(\hat{\pi}_{y,1}-\hat{\pi}_{x,2}\hat{\beta}_{ts2sls}\right)^{T}W_{n}\left(\hat{\beta}_{ts2sls}\right)\left(\hat{\pi}_{y,1}-\hat{\pi}_{x,2}\hat{\beta}_{ts2sls}\right)\stackrel{d}{\rightarrow}\chi_{k_{z}-1}^{2}.
\]
These are the results as in \citet{ZhaoetalStatSci2019}.

\section{Some Simulation Results}

\label{sec:Some-Simulation-Results}

We conduct a Monte Carlo study to simulate the performance of the
proposed two-sample estimators and the two-sample over-identification
test. The DGP follows a two-sample design: sample~1 ($n_{1}=500$)
contains $(Y_{1},Z_{1},c_{1})$ and sample~2 ($n_{2}=1000$) contains
$(X_{2},Z_{2},c_{2})$. We simulate $10{,}000$ replications with
$k_{x}=2$ endogenous regressors and $k_{z}=3$ instruments. The structural
parameters are $(\beta_{1},\beta_{2},\beta_{c},\beta_{0})=(0.3,-0.1,0.1,0.2)$.
Two designs are considered: (i) conditional homoskedasticity and (ii)
conditional heteroskedasticity. This follows the specifications of
simulation study in the supplementary materials of \citet{PaciniWindEL2016}.

\begin{table}[htbp]
\begin{centering}
{\small\caption{{\small} Monte Carlo Results.}\label{tab:MC_1to2}
 }{\small{}%
\begin{tabular}{llcccccc}
\toprule 
{\small Design } & {\small Estimator } & {\small$\hat{\beta}$ } & {\small sd($\hat{\beta}$) } & {\small se$_{\mathrm{hom}}$ } & {\small se$_{\mathrm{het}}$ } & {\small Wald } & {\small Wald$_{\mathrm{rob}}$ }\tabularnewline
\midrule 
\multicolumn{8}{l}{{\small\textbf{Homoskedastic}}}\tabularnewline
$\beta_{1}$ & {\small ts2sls} & {\small 0.2998 } & {\small 0.0745 } & {\small 0.0747 } & {\small 0.0742 } & {\small 0.0478 } & {\small 0.0501 }\tabularnewline
 & {\small rob 2-step} & {\small 0.2998 } & {\small 0.0745 } & {\small -- } & {\small 0.0740 } & {\small -- } & {\small 0.0503 }\tabularnewline
{\small$\beta_{2}$ } & {\small ts2sls} & {\small -0.0991 } & {\small 0.0851 } & {\small 0.0843 } & {\small 0.0836 } & {\small 0.0527 } & {\small 0.0544 }\tabularnewline
 & {\small rob 2-step } & {\small -0.0991 } & {\small 0.0853 } & {\small -- } & {\small 0.0835 } & {\small -- } & {\small 0.0562 }\tabularnewline
\midrule 
\multicolumn{8}{l}{{\small\textbf{Heteroskedastic}}}\tabularnewline
{\small$\beta_{1}$ } & {\small ts2sls} & {\small 0.2999 } & {\small 0.1026 } & {\small 0.0736 } & {\small 0.0995 } & {\small 0.1579 } & {\small 0.0503 }\tabularnewline
 & {\small rob 2-step} & {\small 0.3003 } & {\small 0.0914 } & {\small -- } & {\small 0.0892 } & {\small -- } & {\small 0.0528 }\tabularnewline
{\small$\beta_{2}$} & {\small ts2sls} & {\small -0.0998 } & {\small 0.0991 } & {\small 0.0828 } & {\small 0.0962 } & {\small 0.0951 } & {\small 0.0517 }\tabularnewline
 & {\small rob 2-step } & {\small -0.0999 } & {\small 0.0908 } & {\small -- } & {\small 0.0887 } & {\small -- } & {\small 0.0537 }\tabularnewline
\bottomrule
\end{tabular}}{\small{} }{\small\par}
\par\end{centering}
{\footnotesize\textbf{Notes:}}{\footnotesize{} Mean statistics from
10{,}000 Monte Carlo replications. $n_{1}=500$, $n_{2}=1000$.
Wald and Wald$_{\mathrm{rob}}$ report rejection frequencies at the
5\% nominal level.}{\footnotesize\par}
\end{table}

\begin{table}[htbp]
\begin{centering}
\caption{ Monte Carlo results for the two-sample over-identification test.}\label{tab:MC_OID}
\begin{tabular}{lcccc}
\toprule 
 & \multicolumn{2}{c}{Homoskedastic design} & \multicolumn{2}{c}{Heteroskedastic design}\tabularnewline
\midrule 
Estimator  & OID  & RF  & OID  & RF \tabularnewline
\midrule 
2s2sls  & 1.000  & 0.0508  & 1.735  & 0.1367 \tabularnewline
rob 2-step  & 1.022  & 0.0522  & 1.044  & 0.0528 \tabularnewline
\bottomrule
\end{tabular}
\par\end{centering}
{\footnotesize\textbf{Notes:}}{\footnotesize{} Entries report the mean
over-identification (Sargan/Hansen) test statistic and the associated
rejection frequencies (RF) across 10,000 Monte Carlo replications.
Nominal test size is 5\%. }{\footnotesize\par}
\end{table}

Tables \ref{tab:MC_1to2} and \ref{tab:MC_OID} summarize the main
Monte Carlo findings. Across both designs, all estimators appear unbiased.
Under homoskedasticity, homoskedastic and heteroskedastic standard
errors closely track empirical standard deviations, coverage is near
nominal, and over-identification tests are correctly sized. In contrast,
under heteroskedasticity, homoskedastic standard errors substantially
understate sampling variability, leading to pronounced over-rejection
of Wald and Sargan tests. Heteroskedasticity-robust methods restore correct inference, and the robust two-step estimator typically achieves the smallest empirical dispersion while delivering well-calibrated
Hansen over-identification tests. Overall, the results highlight the
importance of robust weighting for valid inference and specification
testing in two-sample settings.

\section{Weak-Instruments Tests, Bias}

\label{sec:Weak-Instruments-Tests,-Bias}

\subsection{Homoskedasticity and homogeneity}

Under conditional homoskedasticity and homogeneity Assumption \ref{ass:homg}
we consider the standard first-stage $F$-statistic in sample 2, given
by 
\[
F_{\hat{\pi}_{x,2}}=\hat{\pi}_{x,2}^{\prime}\hat{V}_{\hat{\pi}_{x,2}}^{-1}\hat{\pi}_{x,2}/k_{z}.
\]
Under weak-instrument asymptotics, as in \citet{Staiger1997} and
\citet{Stock2005}, given by the representation 
\[
\pi_{x}=c/\sqrt{n_{2}},
\]
we find that the same critical values can be used as tabulated by
\citet{Stock2005} in relation to the weak-instruments test in terms
of relative bias of the ts2sls estimator. Whereas in the one-sample
case this relative bias is relative to the OLS estimator of $\beta$,
in the two-sample case this relative bias is given by the proportional
bias 
\[
PB_{n}=\left|\frac{\mathbb{E}\left[\hat{\beta}_{ts2sls}\right]-\beta}{\beta}\right|.
\]
This is due to the fact that the weak-instruments bias of the ts2sls
estimator is towards $0$. In the limit under the weak-instruments
asymptotics we get 
\[
PB_{n}\rightarrow\left|\mathbb{E}\left[\frac{\left(\lambda+\xi_{2}\right)'\xi_{2}}{\left(\lambda+\xi_{2}\right)'\left(\lambda+\xi_{2}\right)}\right]\right|
\]
where $\lambda=\Sigma_{\hat{\pi}_{x,2}}^{-1/2}c$ and $\xi_{2}\sim\mathcal{N}\left(0,I_{k_{z}}\right)$.
This is the same expression as in \citet{Stock2005} for the 2sls
bias relative to the OLS bias in the one-sample case.

For the $F$-statistic we have the weak-instruments asymptotics result
that 
\[
F_{\hat{\pi}_{x,2}}\stackrel{d}{\rightarrow}\chi_{k_{z}}^{2}\left(\lambda'\lambda\right)/k_{z},
\]
which matches the one-sample case. Hence the same critical values for
$F_{\hat{\pi}_{x,2}}$ for a maximum proportional bias of say 10\%
apply. See the Appendix for further derivations.

\subsection{Effective F-statistic}

Under general forms of heteroskedasticity we can use the results in
\citet{Windmeijer2025} who generalized the use of the effective F-statistic
as proposed by \citet{Olea2013} from testing weak instruments in terms
of the Nagar bias of the 2sls estimator to a general class of linear
GMM estimators. For the ts2sls estimator, the generalization of the
effective F-statistic is given by

\[
\hat{F}_{\text{eff}}\left(\hat{V}_{\hat{\pi}_{y,1}}\right)=\frac{\hat{\pi}_{x,2}^{\prime}\hat{V}_{y,1}^{-1}\hat{\pi}_{x,2}}{\text{tr}\left(\hat{V}_{r,\hat{\pi}_{x,2}}\hat{V}_{\hat{\pi}_{y,1}}^{-1}\right)}.
\]

As shown in the Appendix, the worst-case benchmark bias is again $-\beta$,
and as also shown in the Appendix, the critical value function of
\citet{Olea2013} applies directly.

\section{Application}

\label{sec:Application}

We illustrate the practical relevance of the proposed two-sample estimators
using the education and political affiliation study of \citet{marshall2019anti}.
The setting is canonical for two-sample IV: educational attainment
and voting outcomes are observed in separate surveys over the same
time period but share common instruments. This makes it a suitable environment to assess efficiency, inference, and specification testing
in two-sample designs.

Outcome data are drawn from the National Annenberg Election Survey
(NAES), covering U.S. presidential elections in 2000, 2004, and 2008,
while educational attainment is obtained from the American Community
Survey (ACS) for corresponding cohorts. The samples are linked by
cohort and state of birth and the instruments are state-level compulsory
schooling (drop-out) laws, which vary by cohort and state. The outcomes
measure Democratic party self-identification, voting intention, and
past voting behavior.

We re-estimate Marshall's baseline specifications using the ts2sls
and efficient two-step estimators. Inference is conducted using both
homoskedastic and heteroskedastic (cluster-robust) variance estimators,
and over-identification tests are computed using the corresponding
weighting matrices. 

Estimation and test results are presented in Table \ref{tab:AppliedResults}.
Across all outcomes, we replicate the ts2sls estimates of \citet[Table 2]{marshall2019anti},
who presented cluster-robust (cr) standard errors, where the clustering
is by state. Whereas in our approach we simply use the cr standard
errors from the two linear regression, using for example the ``sandwich''
routine in R, \citet{marshall2019anti} calculates these separately
and directly, based on the residuals $\hat{u}=y_{1}-\hat{x}_{1}\hat{\beta}_{ts2sls}=y_{1}-Z_{1}\hat{\pi}_{x,2}\hat{\beta}_{ts2sls}$.
These are clearly a type of reduced-form residuals. In our approach
we use the least-squares residuals $\hat{v}_{y,1}=y_{1}-Z_{1}\hat{\pi}_{y,1}$,
and, given the cr robust variance estimates of $\hat{\pi}_{y,1}$and
$\hat{\pi}_{x,2}$, we don't need to perform any further direct calculation,
or indeed access the data. It is easily seen the $\hat{u}'\hat{u}\geq\hat{v}_{y,1}^{\prime}\hat{v}_{y,1}$.
The cluster-robust standard errors are also systematically smaller
than Marshall's when constructed using our method and hence based
on reduced-form least-squares residuals. The standard homoskedastic
first-stage F statistics are large (around 40), while cr effective
F-statistics are substantially smaller (around 8.5) and only marginally
larger than the calculated critical values. The Sargan/Hansen test
results indicate some specification problems for the ``Intend'' outcome. 

\begin{table}[ht]
\begin{centering}
{\footnotesize\caption{{\footnotesize} Two-Sample IV Estimates: Replication of \citet[Table 2]{marshall2019anti}}\label{tab:AppliedResults}
 }{\footnotesize{}%
\begin{tabular}{@{}llccccccccc@{}}
\toprule 
{\footnotesize\textbf{Outcome}}{\footnotesize{} } & Estimator & {\footnotesize$\hat{\beta}_{1}$ } & {\footnotesize se$_{\text{hom}}$ } & {\footnotesize se$_{\text{cr}}$ } & {\footnotesize se$_{\text{marshall}}$} & {\footnotesize OID } & {\footnotesize$p$-val } & {\footnotesize$F$ } & {\footnotesize$F_{\text{eff,cr}}$ } & {\footnotesize cv$_{\text{cr}}$ }\tabularnewline
\midrule 
{\footnotesize\textbf{Partisan}}{\footnotesize{} } & {\footnotesize ts2sls } & {\footnotesize -0.1428} & {\footnotesize 0.0638 } & {\footnotesize 0.0903 } & {\footnotesize 0.0959} & {\footnotesize 2.923 } & {\footnotesize 0.087 } & {\footnotesize 41.37 } & {\footnotesize 8.64 } & {\footnotesize 6.71 }\tabularnewline
 & {\footnotesize rob 2-step } & {\footnotesize -0.1534 } & {\footnotesize -- } & {\footnotesize 0.0901 } &  & {\footnotesize 2.726 } & {\footnotesize 0.099 } & {\footnotesize -- } & {\footnotesize -- } & {\footnotesize -- }\tabularnewline
\midrule 
{\footnotesize\textbf{Intend}}{\footnotesize{} } & {\footnotesize ts2sls} & {\footnotesize -0.1949 } & {\footnotesize 0.0731 } & {\footnotesize 0.0889 } & {\footnotesize 0.0906} & {\footnotesize 0.648 } & {\footnotesize 0.421 } & {\footnotesize 42.37 } & {\footnotesize 8.58 } & {\footnotesize 6.35 }\tabularnewline
 & {\footnotesize rob 2-step } & {\footnotesize -0.1987 } & {\footnotesize -- } & {\footnotesize 0.0887 } &  & {\footnotesize 0.679 } & {\footnotesize 0.410 } & {\footnotesize -- } & {\footnotesize -- } & {\footnotesize -- }\tabularnewline
\midrule 
{\footnotesize\textbf{Voted}}{\footnotesize{} } & {\footnotesize ts2sls} & {\footnotesize -0.1482 } & {\footnotesize 0.0789 } & {\footnotesize 0.0732 } & {\footnotesize 0.0804} & {\footnotesize 4.051 } & {\footnotesize 0.044 } & {\footnotesize 41.37 } & {\footnotesize 8.48 } & {\footnotesize 6.88 }\tabularnewline
 & {\footnotesize rob 2-step } & {\footnotesize -0.1445 } & {\footnotesize -- } & {\footnotesize 0.0732 } &  & {\footnotesize 3.404 } & {\footnotesize 0.065 } & {\footnotesize -- } & {\footnotesize -- } & {\footnotesize -- }\tabularnewline
\bottomrule
\end{tabular}}{\footnotesize\par}
\par\end{centering}
{\footnotesize\textbf{Notes:}}{\footnotesize{} robust estimator and
test statistics are clustering robust (cr)}{\footnotesize\par}
\end{table}

\bibliographystyle{ecta}
\bibliography{2SIV}

\section*{Appendix}

\global\long\def\thesection{A}%
\global\long\def\theequation{A.\arabic{equation}}%
\global\long\def\thetable{A\arabic{table}}%
\setcounter{table}{0}\setcounter{equation}{0}

\subsection{Weak-Instruments F-test}

\subsubsection{Homoskedasticity and Homogeneity, Standard First-Stage $F$-statistic}

We assume homoskedasticity and homogeneity Assumption \ref{ass:homg}.

The standard first-stage $F$-statistic in sample 2 is given by 
\[
F_{\hat{\pi}_{x,2}}=\hat{\pi}_{x,2}^{\prime}\hat{V}_{\hat{\pi}_{x,2}}^{-1}\hat{\pi}_{x,2}/k_{z}.
\]

Under the restriction that $\pi_{y}=\pi_{x}\beta$ and weak-instruments
asymptotics, 
\[
\pi_{x}=\frac{c}{\sqrt{n_{2}}},
\]
we have the limiting distribution results 
\begin{align*}
\sqrt{n_{2}}\hat{\pi}_{y,1} & \stackrel{d}{\rightarrow}\psi_{y,1}\sim\mathcal{N}\left(\beta c,\alpha^{-1}\Sigma_{\hat{\pi}_{y,1}}\right);\\
\sqrt{n_{2}}\hat{\pi}_{x,2} & \stackrel{d}{\rightarrow}\psi_{x,2}\sim\mathcal{N}\left(c,\Sigma_{\hat{\pi}_{x,2}}\right),
\end{align*}
for $n_{1}\rightarrow\infty$, $n_{2}\rightarrow\infty$, with $\frac{n_{1}}{n_{2}}\rightarrow\alpha>0$.

Under homoskedasticity we have that 
\[
\Sigma_{\hat{\pi}_{y,1}}=\sigma_{v_{y,1}}^{2}Q_{z_{1}z_{1}}^{-1};\,\Sigma_{\hat{\pi}_{x,2}}=\sigma_{v_{x,2}}^{2}Q_{z_{2}z_{2}}^{-1},
\]
where $Q_{Z_{1}Z_{1}}=\mathbb{E}\left[z_{1i}z_{1i}^{\prime}\right]$
and $Q_{Z_{2}Z_{2}}=\mathbb{E}\left[z_{2j}z_{2j}^{\prime}\right]$.
Further under homogeneity assumption \ref{ass:homg} it follows that
\[
\Sigma_{\hat{\pi}_{y,1}}=\frac{\sigma_{v_{y,1}}^{2}}{\sigma_{v_{x,2}}^{2}}\Sigma_{\hat{\pi}_{x,2}}.
\]

For the weak-instruments bias of the ts2sls estimator, we then get

\begin{align*}
\hat{\beta}_{ts2sls}-\beta & =\frac{\hat{\pi}_{x,2}^{\prime}\hat{V}_{\hat{\pi}_{y,1}}^{-1}\left(\hat{\pi}_{y,1}-\hat{\pi}_{x,2}\beta\right)}{\hat{\pi}_{x,2}^{\prime}\hat{V}_{\hat{\pi}_{y,1}}^{-1}\hat{\pi}_{x,2}}\\
 & =\frac{\sqrt{n_{2}}\hat{\pi}_{x,2}^{\prime}\left(n_{1}\hat{V}_{\hat{\pi}_{y,1}}\right){}^{-1}\sqrt{n_{2}}\left(\hat{\pi}_{y,1}-\hat{\pi}_{x,2}\beta\right)}{\sqrt{n_{2}}\hat{\pi}_{x,2}^{T}\left(n_{1}\hat{V}_{\hat{\pi}_{y,1}}\right){}^{-1}\sqrt{n_{2}}\hat{\pi}_{x,2}}\\
 & \stackrel{d}{\rightarrow}\frac{\psi_{x,2}^{\prime}\Sigma_{\hat{\pi}_{y,1}}^{-1}\left(\psi_{y,1}-\psi_{x,2}\beta\right)}{\psi_{x,2}^{\prime}\Sigma_{\hat{\pi}_{y,1}}^{-1}\psi_{x,2}}\\
 & =\frac{\psi_{x,2}^{\prime}\Sigma_{\hat{\pi}_{x,2}}^{-1}\left(\psi_{y,1}-\psi_{x,2}\beta\right)}{\psi_{x,2}^{\prime}\Sigma_{\hat{\pi}_{x,2}}^{-1}\psi_{x,2}}\\
 & =\frac{\gamma_{2}^{\prime}\left(\gamma_{1}-\gamma_{2}\beta\right)}{\gamma_{2}^{\prime}\gamma_{2}}
\end{align*}
with 
\begin{align*}
\gamma_{1} & =\Sigma_{\hat{\pi}_{x,2}}^{-1/2}\psi_{y,1}\\
\gamma_{2} & =\Sigma_{\hat{\pi}_{x,2}}^{-1/2}\psi_{x,2}.
\end{align*}
It follows that 
\[
\left(\begin{array}{c}
\gamma_{1}\\
\gamma_{2}
\end{array}\right)\sim\mathcal{N}\left(\left(\begin{array}{c}
\lambda\beta\\
\lambda
\end{array}\right),\left(\begin{array}{cc}
\alpha^{-1}\frac{\sigma_{v_{y,1}}^{2}}{\sigma_{v_{x},2}^{2}}I_{k_{z}} & 0\\
0 & I_{k_{z}}
\end{array}\right)\right),
\]
where $\lambda=\Sigma_{\hat{\pi}_{x,2}}^{-1/2}c$ .

Write 
\begin{align*}
\gamma_{1} & =\beta\lambda+\alpha^{-1/2}\frac{\sigma_{v_{y,1}}}{\sigma_{v_{x,2}}}\xi_{1}\\
\gamma_{2} & =\lambda+\xi_{2},
\end{align*}
where 
\begin{align*}
\xi_{1} & \sim\mathcal{N}\left(0,I_{k_{z}}\right),\\
\xi_{2} & \sim\mathcal{N}\left(0,I_{k_{z}}\right),
\end{align*}
and $\xi_{1}$ and $\xi_{2}$ are independent. Then, we have 
\begin{align*}
\beta^{*} & \coloneqq\frac{\gamma_{2}^{\prime}\left(\gamma_{1}-\gamma_{2}\beta\right)}{\gamma_{2}^{\prime}\gamma_{2}}\\
 & =\frac{\left(\lambda+\xi_{2}\right)'\left(\alpha^{-1/2}\frac{\sigma_{v_{y,1}}}{\sigma_{v_{x,2}}}\xi_{1}-\beta\xi_{2}\right)}{\left(\lambda+\xi_{2}\right)'\left(\lambda+\xi_{2}\right)}\\
 & =-\beta\frac{\left(\lambda+\xi_{2}\right)'\xi_{2}}{\left(\lambda+\xi_{2}\right)'\left(\lambda+\xi_{2}\right)}+\alpha^{-1/2}\frac{\sigma_{v_{y,1}}}{\sigma_{v_{x,2}}}\frac{\left(\lambda+\xi_{2}\right)'\xi_{1}}{\left(\lambda+\xi_{2}\right)'\left(\lambda+\xi_{2}\right)}
\end{align*}

\begin{align*}
\mathbb{E}\left[\beta_{\Omega}^{*}\right] & =-\beta\mathbb{E}\left[\frac{\left(\lambda+\xi_{2}\right)'\xi_{2}}{\left(\lambda+\xi_{2}\right)'\left(\lambda+\xi_{2}\right)}\right]+\alpha^{-1/2}\frac{\sigma_{v_{y,1}}}{\sigma_{v_{x,2}}}\mathbb{E}\left[\frac{\left(\lambda+\xi_{2}\right)'\xi_{1}}{\left(\lambda+\xi_{2}\right)'\left(\lambda+\xi_{2}\right)}\right]\\
 & =-\beta\mathbb{E}\left[\frac{\left(\lambda+\xi_{2}\right)'\xi_{2}}{\left(\lambda+\xi_{2}\right)'\left(\lambda+\xi_{2}\right)}\right]
\end{align*}
as $\xi_{1}$ and $\xi_{2}$ are independent and $\mathbb{E}\left[\xi_{1}\right]=0$.

As a measure of relative bias, we propose 
\[
PB_{n}^{2}=\left(\frac{\mathbb{E}\left[\hat{\beta}_{ts2sls}\right]-\beta}{\beta}\right)^{2}\rightarrow\left(\frac{\mathbb{E}\left[\beta_{\Omega}^{*}\right]}{\beta}\right)^{2},
\]
as $n_{1},n_{2}\rightarrow\infty$, or for the absolute percentage
bias of the ts2sls estimator, 
\[
PB_{n}=\left|\frac{\mathbb{E}\left[\hat{\beta}_{ts2sls}\right]-\beta}{\beta}\right|\rightarrow\left|\frac{\mathbb{E}\left[\beta_{\Omega}^{*}\right]}{\beta}\right|=\left|\mathbb{E}\left[\frac{\left(\lambda+\xi_{2}\right)'\xi_{2}}{\left(\lambda+\xi_{2}\right)'\left(\lambda+\xi_{2}\right)}\right]\right|,\quad\text{for}\:\beta\neq0.
\]

When $\beta=0$, the weak-instruments asymptotics bias of the ts2sls
estimator is $0$.

\subsubsection{Heteroskedasticity and/or Heterogeneity, the Generalized Effective
$F$-Statistic}

As above, we have under weak-instruments asymptotics,

\begin{align*}
\hat{\beta}_{ts2sls}-\beta & =\frac{\hat{\pi}_{x,2}^{\prime}\hat{V}_{\hat{\pi}_{y,1}}^{-1}\left(\hat{\pi}_{y,1}-\beta\hat{\pi}_{x,2}\right)}{\hat{\pi}_{x,2}^{\prime}\hat{V}_{\hat{\pi}_{y,1}}^{-1}\hat{\pi}_{x,2}}\\
 & \stackrel{d}{\rightarrow}\beta^{*}\coloneqq\left(\gamma_{2}^{\prime}\gamma_{2}\right)^{-1}\left(\gamma_{2}^{\prime}\left(\gamma_{1}-\beta\gamma_{2}\right)\right)
\end{align*}
with 
\begin{align*}
\gamma_{1} & =\Omega^{1/2}\psi_{y,1}\\
\gamma_{2} & =\Omega^{1/2}\psi_{x,2},
\end{align*}
with 
\[
n_{1}\hat{V}_{\hat{\pi}_{y,1}}^{-1}\stackrel{p}{\rightarrow}\Omega.
\]
Therefore 
\[
\left(\begin{array}{c}
\gamma_{1}\\
\gamma_{2}
\end{array}\right)\sim\mathcal{N}\left(\left(\begin{array}{c}
\kappa\beta\\
\kappa
\end{array}\right),\left(\begin{array}{cc}
V_{1} & 0\\
0 & V_{2}
\end{array}\right)\right),
\]
where $\kappa=\Omega c$, $V_{1}=\alpha^{-1}\Omega^{1/2}\Sigma_{\hat{\pi}_{y,1}}\Omega^{1/2}$
and $V_{2}=\Omega^{1/2}\Sigma_{\hat{\pi}_{x,2}}\Omega^{1/2}$.

Let 
\[
\hat{F}_{\text{geff}}\left(\hat{V}_{\hat{\pi}_{y,1}}^{-1}\right)=\frac{\hat{\pi}_{x,2}^{\prime}\hat{V}_{\hat{\pi}_{y,1}}^{-1}\hat{\pi}_{x,2}}{\text{tr}\left(\hat{V}_{r,\hat{\pi}_{x,2}}\hat{V}_{\hat{\pi}_{y,1}}^{-1}\right)}\stackrel{d}{\rightarrow}\frac{\gamma_{2}^{\prime}\gamma_{2}}{\text{tr}\left(V_{2}\right)}.
\]
Let 
\begin{align*}
S_{1} & \coloneqq\text{var}\left(\gamma_{1}-\beta\gamma_{2}\right)=V_{1}+\beta^{2}V_{2}\\
S_{12} & \coloneqq\text{cov}\left(\gamma_{2},\gamma_{1}-\beta\gamma_{2}\right)=-\beta V_{2}
\end{align*}
and 
\begin{align*}
\xi & \coloneqq S_{1}^{-1/2}\left(\gamma_{1}-\beta\gamma_{2}\right)\sim\mathcal{N}\left(0,I_{k_{z}}\right)\\
\nu & \coloneqq V_{2}^{-1/2}\left(\gamma_{2}-\kappa\right)\sim\mathcal{N}\left(0,I_{k_{z}}\right)
\end{align*}
we then have that 
\begin{align*}
\beta^{*} & =\frac{\gamma_{2}^{\prime}\left(\gamma_{1}-\beta\gamma_{2}\right)}{\gamma_{2}^{T}\gamma_{2}}\\
 & =\frac{\kappa^{'}S_{1}^{1/2}\xi+\nu^{'}V_{2}^{1/2}S_{1}^{1/2}\xi}{\kappa'\kappa+2\kappa'V_{1}^{1/2}\nu+\nu'V_{2}\nu}.
\end{align*}
Let $\kappa_{0}:=\kappa/\Vert\kappa\Vert$, with $\Vert\kappa\Vert=\sqrt{\kappa'\kappa}$,
and $\mu^{2}:=\Vert\kappa\Vert^{2}/\text{tr}\left(V_{2}\right)$.
It then follows that 
\[
\Vert\kappa\Vert\beta_{\Omega}^{*}=\frac{\kappa_{0}^{\prime}S_{1}^{1/2}\xi+\frac{\nu'V_{2}^{1/2}S_{1}^{1/2}\xi}{\sqrt{\text{tr}\left(V_{2}\right)}}\mu^{-1}}{1+\frac{2\kappa_{0}^{\prime}V_{2}^{1/2}\nu}{\sqrt{\text{tr}\left(V_{2}\right)}}\mu^{-1}+\frac{\nu'V_{2}\nu}{\text{tr}\left(V_{2}\right)}\mu^{-2}}.
\]
Then, from Rothenberg (1984, (6.2)), we get the second-order Edgeworth,
Nagar (1959) approximation 
\begin{align*}
\mathbb{E}\left[\beta^{*}\right] & \approx\frac{1}{\mu^{2}}\frac{1}{\text{tr}\left(V_{2}\right)}\mathbb{E}\left[\nu'V_{2}^{1/2}S_{1}^{1/2}\xi-2\kappa_{0}'S_{1}^{1/2}\xi\kappa_{0}V_{2}^{1/2}\nu\right]\\
 & =\frac{1}{\mu^{2}}\left(\frac{\text{tr}\left(S_{12}\right)-2\kappa_{0}^{\prime}S_{12}\kappa_{0}}{\text{tr}\left(V_{2}\right)}\right)\\
 & =-\frac{\beta}{\mu^{2}}\left(1-\frac{2\kappa_{0}^{\prime}V_{2}\kappa_{0}}{\text{tr}\left(V_{2}\right)}\right)
\end{align*}
Worst case, benchmark, bias, 
\begin{align*}
BM & \approx\frac{\mathbb{E}\left[\gamma_{2}^{\prime}\left(\gamma_{1}-\beta\gamma_{2}\right)\right]}{\mathbb{E}\left[\gamma_{2}^{\prime}\gamma_{2}\right]}=\frac{-\beta\text{tr}\left(V_{2}\right)}{\text{tr}\left(V_{2}\right)\left(1+\mu^{2}\right)}\\
 & =-\frac{\beta}{1+\mu^{2}}.
\end{align*}
So the worst case bias is $-\beta$ when $\mu^{2}=0$.

Let 
\begin{align*}
B\left(V_{2}\right) & \coloneqq\sup_{\kappa_{0}}\left(1-\frac{2\kappa_{0}^{\prime}V_{2}\kappa_{0}}{\text{tr}\left(V_{2}\right)}\right)\\
 & =1-\frac{2\lambda_{\text{min}}\left(V_{2}\right)}{\text{tr}\left(V_{2}\right)}\leq1,
\end{align*}
where $\lambda_{\text{min}}\left(V_{2}\right)$ is the minimum eigenvalue
of $V_{2}$.

The null hypothesis of weak instruments is specified as 
\[
H_{0}:\mu^{2}\in\mathcal{H}\left(V_{2},\tau\right)\,\,\,\text{against}\,\,\,H_{1}:\mu^{2}\ensuremath{\notin}\mathcal{H}\left(V_{2},\tau\right),
\]
where 
\[
\mathcal{H}\left(V_{2},\tau\right)=\left\{ \mu^{2}\in\mathbb{R}_{+}:\mu^{2}<\frac{B\left(V_{2}\right)}{\tau}\right\} .
\]

The test for weak instruments is then based on $\hat{F}_{\text{geff}}\left(\hat{V}_{\hat{\pi}_{y,1}}^{-1}\right)$
which is asymptotically distributed as $\gamma_{2}^{T}\gamma_{2}/\text{tr}\left(V_{2}\right)$
which has mean $1+\mu^{2}$. It follows that we reject $H_{0}$ when
$\hat{F}_{\text{geff}}\left(\hat{V}_{\hat{\pi}_{y,1}}^{-1}\right)$
is large. Denote by $F_{\kappa,V_{2}}^{-1}\left(\alpha\right)$ the
upper $\alpha$ quantile of the distribution of $\gamma_{2}^{T}\gamma_{2}/\text{tr}\left(V_{2}\right)$
and let 
\[
cv\left(\alpha,V_{2},d\right):=\sup_{\kappa\in\mathbb{R}^{k_{z}}}\left\{ F_{\kappa,V_{2}}^{-1}\left(\alpha\right)1_{\left(\frac{\kappa'\kappa}{\text{tr}\left(V_{2}\right)}<d\right)}\right\} ,
\]
where $1_{\left(A\right)}$ denotes the indicator function over a
set $A$. Let 
\[
\hat{V}_{2}=\hat{V}_{\hat{\pi}_{y,1}}^{-1/2}\hat{V}_{r,\hat{\pi}_{x,2}}\hat{V}_{\hat{\pi}_{y,1}}^{-1/2}.
\]
The null of weak instruments is then rejected if 
\[
\hat{F}_{\text{geff}}\left(\hat{V}_{\hat{\pi}_{y,1}}^{-1}\right)>cv\left(\alpha,\hat{V}_{,2},B\left(\hat{V}_{2}\right)/\tau\right).
\]

The critical values can be obtained by Monte Carlo methods, or by
the Patnaik (1949) curve-fitting methodology. The Patnaik critical
value is obtained as the the upper $\alpha$ quantile of $\chi_{\hat{k}_{\text{geff}}\left(\hat{V}_{\hat{\pi}_{y,1}}^{-1}\right)}^{2}\left(d_{\Omega_{n},\tau}\hat{k}_{\text{geff}}\left(\hat{V}_{\hat{\pi}_{y,1}}^{-1}\right)\right)/\hat{k}_{\text{geff}}\left(\hat{V}_{\hat{\pi}_{y,1}}^{-1}\right)$
where $\chi_{\hat{k}_{\text{geff}}\left(\hat{V}_{\hat{\pi}_{y,1}}^{-1}\right)}^{2}\left(d_{\tau}\hat{k}_{\text{geff}}\left(\hat{V}_{\hat{\pi}_{y,1}}^{-1}\right)\right)$
denotes the noncentral $\chi^{2}$ distribution with $\hat{k}_{\text{geff}}\left(\hat{V}_{\hat{\pi}_{y,1}}^{-1}\right)$
degrees of freedom and noncentrality parameter $d_{\tau}\hat{k}_{\text{geff}}\left(\hat{V}_{\hat{\pi}_{y,1}}^{-1}\right)$,
with

\begin{align}
d_{\tau} & =B\left(\hat{V}_{2}\right)/\tau;\label{eq:dOmn}\\
\hat{k}_{\text{geff}}\left(\hat{V}_{\hat{\pi}_{y,1}}^{-1}\right) & =\frac{\left[\text{tr}\left(\hat{V}_{2}\right)\right]^{2}\left(1+2d_{\tau}\right)}{\text{tr}\left(\hat{V}_{2}^{\prime}\hat{V}_{2}\right)+2d_{\Omega_{n},\tau}\text{tr}\left(\hat{V}_{2}\right)\lambda_{\text{max}}\left(\hat{V}_{2}\right)},\label{eq:keff}
\end{align}
and where $\lambda_{\text{max}}\left(\hat{V}_{2}\right)$ denotes
the maximum eigenvalue of $\hat{V}_{2}$.

Under homoskedasticity, we get 
\[
\hat{F}_{\text{eff}}\left(\hat{V}_{y,1}\right)=\frac{\hat{\pi}_{x,2}^{T}\hat{V}_{\hat{\pi}_{y,1}}^{-1}\hat{\pi}_{x,2}}{\text{tr}\left(\hat{V}_{\hat{\pi}_{x,2}}\hat{V}_{\hat{\pi}_{y,1}}^{-1}\right)},
\]
which becomes the standard $F$-statistic under homogeneity.
\end{document}